\documentclass[conference]{IEEEtran}
\IEEEoverridecommandlockouts
\usepackage{cite}
\usepackage{amsmath,amssymb,amsfonts}
\usepackage{algorithmic}
\usepackage{graphicx}
\usepackage{textcomp}
\usepackage{xcolor}
\PassOptionsToPackage{hyphens}{url}
\usepackage{tabularx}
\usepackage{subcaption}
\usepackage{booktabs}

\newif\ifshowcomments
\showcommentstrue

\ifshowcomments
\newcommand{\mynote}[2]{\textcolor{blue}{\fbox{\bfseries\sffamily\scriptsize#1}}
  \textcolor{blue}{{$/*$\textsf{\emph{#2}}$*/$}}}
\else
\newcommand{\mynote}[2]{}
\fi

\def\BibTeX{{\rm B\kern-.05em{\sc i\kern-.025em b}\kern-.08em
    T\kern-.1667em\lower.7ex\hbox{E}\kern-.125emX}}

\begin{document}

\title{The Separator, a Two-Phase Oil and Water Gravity CPS Separator Testbed}

\author{\IEEEauthorblockN{Michael Breza\IEEEauthorrefmark{1},
Laksh Bhatia\IEEEauthorrefmark{1}, 
Ivana Tomi\'c\IEEEauthorrefmark{2},
Anqi Fu\IEEEauthorrefmark{1},  
Waqas Ikram\IEEEauthorrefmark{4}, Valentinos Kongezos\IEEEauthorrefmark{4},
\\ Julie A. McCann\IEEEauthorrefmark{1}}
\IEEEauthorblockA{\IEEEauthorrefmark{1}Department of Computing,
Imperial College London, UK\\
\IEEEauthorrefmark{2}Department of Computing and Information Systems, University of Greenwich, London, UK\\
\IEEEauthorrefmark{4} ABB, Oslo, Norway\\
Email: michael.breza04@imperial.ac.uk, laksh.bhatia16@imperial.ac.uk, i.tomic@gre.ac.uk, \\ anqifu@ieee.org, waqas.ikram@no.abb.com, valentinos.kongezos@no.abb.com, \\ j.mccann@imperial.ac.uk}}

\maketitle

\begin{abstract}
Industrial Control Systems (ICS) are evolving with advances in new technology. The addition of wireless sensors and actuators and new control techniques means that engineering practices from communication systems  
are being integrated into those used for control systems. The two are engineered in very different ways. 
Neither engineering approach is capable of accounting for the subtle interactions and interdependence that occur 
when the two are combined. This paper describes our first steps to bridge this gap, and push the boundaries of both
computer communication system and control system design. We present The Separator testbed, a Cyber-Physical testbed enabling our search for a suitable way to engineer systems that combine both computer networks and control 
systems. 
\end{abstract}


\section{Introduction}

Industrial Control Systems (ICS) are used in civil infrastructure (water and gas distribution networks, power grids, transportation) and industrial applications
(process plants, automotive industry). 
The automation of these systems is constantly
evolving with new technologies to increase their efficiency, safety and reduce human management effort.
One area currently changing ICS is the integration of wireless communication systems. The
inclusion of new technologies creates new interactions between system components that previously did not exist. 
This paper presents a new ICS testbed that enables us to explore these new interactions.

WirelessHART and ISA 100.11a\cite{6102417} are wireless communication protocols used for ICS 
applications. They were designed for traditional sensing applications, standardised to be 
used in industry and are used without any consideration for the influences that the
control application has on them, and the affects that they have on the control application. 
For instance, if network interference slows down the data rate of the sensors to the 
controller, the controller may not have the up-to-date information that it needs to maintain 
system stability. On the other hand, if the controller is using a sampling scheme that can 
vary based on system state, then a disturbance to the system may cause the sensors to
send data at a rate higher than the network channel capacity.

This notion of coupling between the control system, communication system and computation is 
the core idea of Cyber-Physical Systems (CPS). The aim of CPS is to develop approaches to 
engineer these coupled system components in a way that ensures the entire system meets its 
design goals of stability and safety.




A common approach to study the interactions between system components such as communication and control 
is through mathematical models and simulations. 
Control systems are modelled analytically using differential equations, computer systems are modelled using simulations 
and discrete mathematical representations and these models do not combine easily\cite{5995279}. There exist 
simulation environments such as Ptolemy II \cite{ptolemaeus2014system}, but they use a high level of 
abstraction and need a special notion of time to integrate the continuous and 
discrete domains. The high level of abstraction obscures realistic effects like 
radio interference, temperature effects on the physical phenomenon, etc. These effects are difficult to 
categorise and hard to include in simulation, yet they may have a large impact when these 
systems are deployed in the wild. 

In this paper we present a CPS simulation testbed called The Separator.
It overcomes the loss of accuracy due to abstraction associated with the use of simulations 
and models. The Separator allows us to reliably emulate the behaviour of a real 
system and accurately provide the physical characteristics of sensors and actuators as well 
as the unpredictability of wireless communication systems. It also gives us the ability to 
perform repeatable experiments so that we can categorise and understand the nature of CPS 
interactions present in the system. 

\label{specifications}
We collaborated with ABB to address the questions of the design, engineering and safety of a 
testbed suitable to explore the CPS interactions in ICS. They are keenly aware of the 
challenges, risks and issues associated with ICS engineering. Together, we developed the 
following testbed requirements.

\begin{enumerate}
    \item Have a real physical process that lends itself to control and has a well defined notion of stability and safety.
    \item Use industrial grade sensors and actuators for sensing and actuation of the physical process to  approximate the behaviour on a real ICS.  Domestic components can not handle the demands of industrial 
    applications, such as pressure for valves, current for wiring or to be operated constantly for long periods of 
    time.  
    \item Control the process with an industrial grade controller that has the same performance as one that would be used on a production system. 
    \item Use state-of-art wireless network connecting the sensors to the controller to allow us to observe the performance of the system with different radio environments. 
    \item Build a system that can be used to address the issues of communication and control co-design and capture the subtle interactions between sensing, communication, control and computation. 
    \item Design a safe system that can be used by students and researchers in an academic environment.
\end{enumerate}

ABB addressed all of the diverse engineering challenges that are associated 
with the development of a CPS testbed with a real physical process. Together 
we developed \textit{The Separator, a two phase, oil and water gravity separator with senors communicating using WirelessHART that uses industrial grade components}.

In this paper we review previous approaches to the engineering of ICS systems, present the specifications of The Separator, show examples of the type of research that it is capable of,briefly describe what we have learned so far from its specification and building, and conclude with a description of the research that we intend to pursue in the future.

\section{Related Work}


CPS interactions can be studied with two methods. One is to use models and simulation tools and second by using tesbteds. We 
first discuss some of the CPS simulators and then describe some CPS testbeds. For a more in-depth discussion please refer to 
\cite{8464069}.

There are model based approaches that use computer systems to create models of CPSs. Systems such as
Ptolemy II \cite{ptolemaeus2014system} or GISOO \cite{6700049} combine discrete state-based models of
computer systems with continuous models of physical systems and their controllers. 
Modelling platforms do not use input data from real sensors and actuators that contain noise
from the physical sensing process and radio environment. This kind of noise is difficult to
correctly categorise and include in simulation, yet can have a profound impact on the system 
when it is deployed in the wild.


There are a number of CPS testbeds simulating smart grids and water distribution networks.
The Secure Water Treatment (SWaT) testbed\cite{mathur2016swat} was designed to assess the security 
vulnerabilities of water treatment plants. 
The Water Distrubition Tesbted (WADI) \cite{Ahmed:2017:WWD:3055366.3055375} is used for detecting cyber-attacks and physical 
attacks. The INVITED\cite{8060334} testbed is designed to test the timing behaviours of CPSs.
In \cite{morris2011control}, the authors have created a SCADA testbed with a focus on security research. This testbed uses industrial grade sensors and controller but does not use industrial wireless protocols. In \cite{Araujo2014}, the authors use 
various aperiodic control schemes with 802.15.4 to control a double-tank system. The above testbeds have been designed for 
security research or timing analysis but none of them were designed to study CPS interactions nor do they use state-of-the-art communication protocols designed for control like WirelessHART. 
Other testbeds \cite{Hernandez504854} focus on systems with pendulums as the physical process. Although Pendulums are an 
accepted benchmark for the control theory community \cite{boubaker2012inverted}, our industrial partners suggested two-phase oil and water separation as a physical process that is more representative in the process industry.



The CPS testbed that is closest to our requirements is the WaterBox\cite{Kartakis:2015:WTM:2738935.2738939}. It 
was designed with domestic components, a controllable process and used 802.11 wireless communication at the time of publication. The small-scale domestic controllable valves on
the WaterBox can only handle a small total line pressure, and adjust themselves from open to 
close in $1$ second. We wanted industrial grade controllable valves and sensors. 
The one used in our testbed is designed for higher line pressures, but requires $9$ seconds to adjust 
themselves from fully open to fully closed. These minor differences are very important to the fidelity of our 
testbed. 


In the next section we describe the physical and digital architecture of The Separator.

\section{The Separator Architecture}
The Separator testbed is the result of the design requirements given in Section~\ref{specifications}. A summary of
the Separator physical process is given next, followed by a more in-depth description of the Separator's design overview
and the components and the way that each of our requirements has been met. Finally, we present
the safety features of The Separator. 

\subsection{The Separator Physical Process}
Control of liquid levels in tanks and flows between tanks are basic problems in process industry\cite{8245161}. And so, the physical process that we use is oil and water gravity separation. It is used in the petroleum industry. When the petroleum mixture is extracted, the oil is mixed with water and other impurities. This mixture is put into a tank 
where the water settles to the bottom, and the oil floats on top at the rate of separation. The placement of a 
simple barrier in the middle of the tank allows one part of the tank to have only water at the bottom, and the 
other to have only oil. The water and oil can then be separated and drained into individual tanks by putting 
automatic valves at the bottom of the tank. The level of oil and water 
in the separation tank can be kept constant by controlling the degree to which the valves are open. 

The oil separation process lends itself to a clear definition of stability. The oil and the water levels, in their 
respective sections of the separation tank, are set by an operator. The controller maintains the oil and water 
levels by setting the opening degree of the valves. The degree over or under the set point of the liquid levels is 
referred to as the overshoot, or undershoot. We can use the maximum size of the overshoot above the set point as 
the measure of system stability. 

The notion of safety is defined in terms of the stability. If the total liquid level in the tank with its overshoot are below a certain level, we can say that the system is operating safely. If the total liquid level with its overshoot exceeds 80\% of the total capacity of 
the tank, or if the water level exceeds its section and enters the oil section, we say that the system is operating in an unsafe way. 
This clean distinction between safe and unsafe states is common to all ICS applications, and gives us a clear set of parameters to use for analysis.

These notions will be made clear in the next section when we present the actual architecture of The Separator.


%
%

\subsection{The Separator Design Overview}

\begin{figure}
    \centering
    \includegraphics[width=\linewidth]{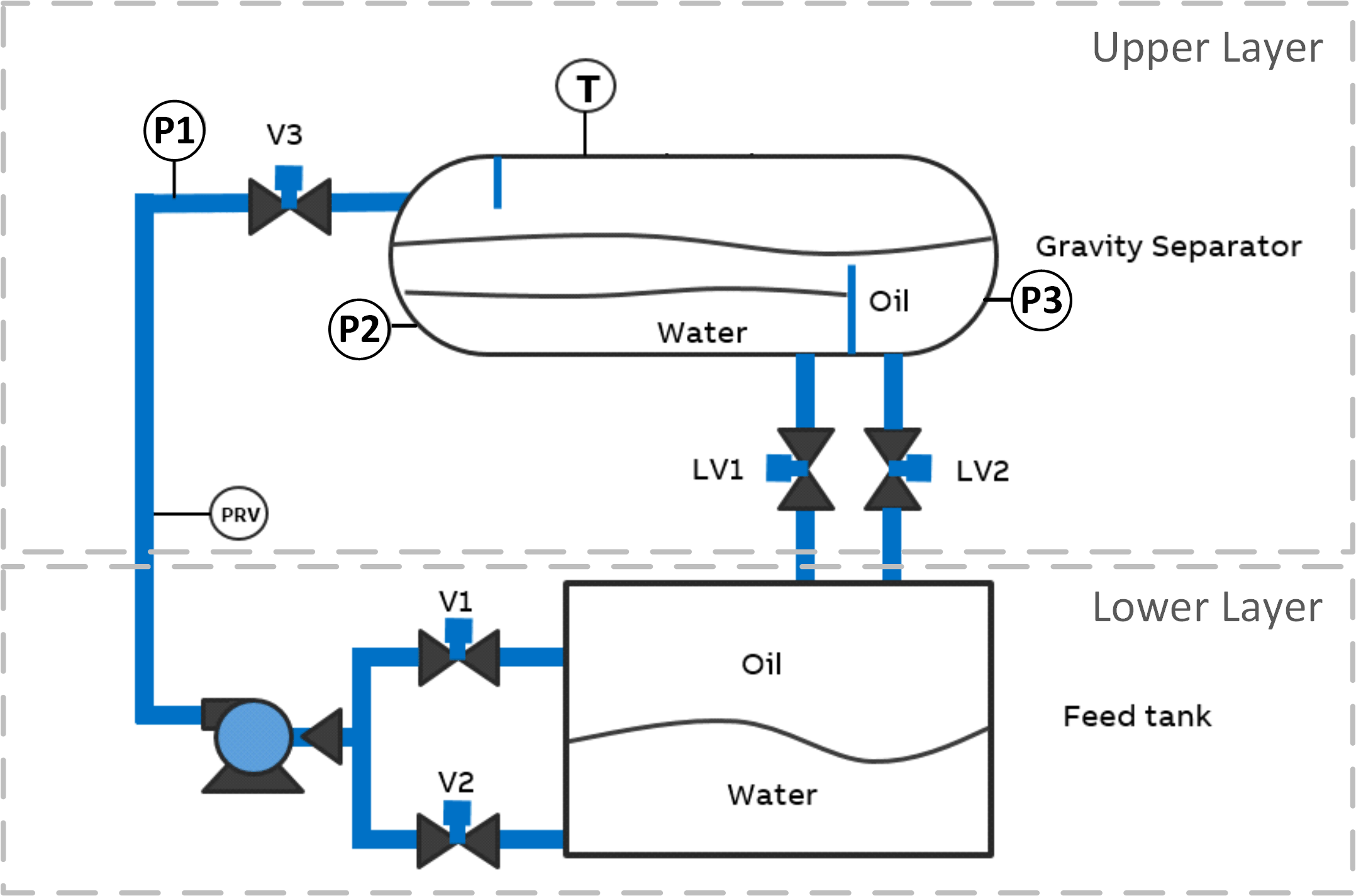}
    \caption{The Separator Design Overview}
    \vspace{-5mm}
    \label{fig:separator}
\end{figure}

The design overview of The Separator is depicted in Figure~\ref{fig:separator}.
The Separator consists of two individual layers:\vspace{2pt}\\
\textbf{Lower layer} - A feed tank holding $100$ litres of ionised water and $105$ liters of Exxsol D-60 oil. 
The tank has two feed valves (V1 and V2), one for the water and one for the
oil. An electrical impeller pump mixes and pumps the oil and water to the upper layer.\vspace{2pt}\\
\textbf{Upper layer} - A $60$ litre separation tank in which the separation process occurs. The tank is
fitted with wireless sensors (P1, P2, P3 and T) and wired actuators (LV1 and LV2). The oil and water mixture is pumped 
into the separation tank by the pump mentioned above via inlet valve (V3). The separation tank
is divided into two sections by a separation plate. The left section holds water and oil and is where gravity
separation occurs. The right section receives the overflow from the left section, and contains only oil.
There are valves at the bottom of the separation tank, one in the left section which only drains water (LV1),
and one in the right section for oil (LV2). The water and oil levels are regulated by the PID controller which operates 
the valves.



\subsection{The Separator Components}

\begin{figure}
    \centering
    \includegraphics[width=\linewidth]{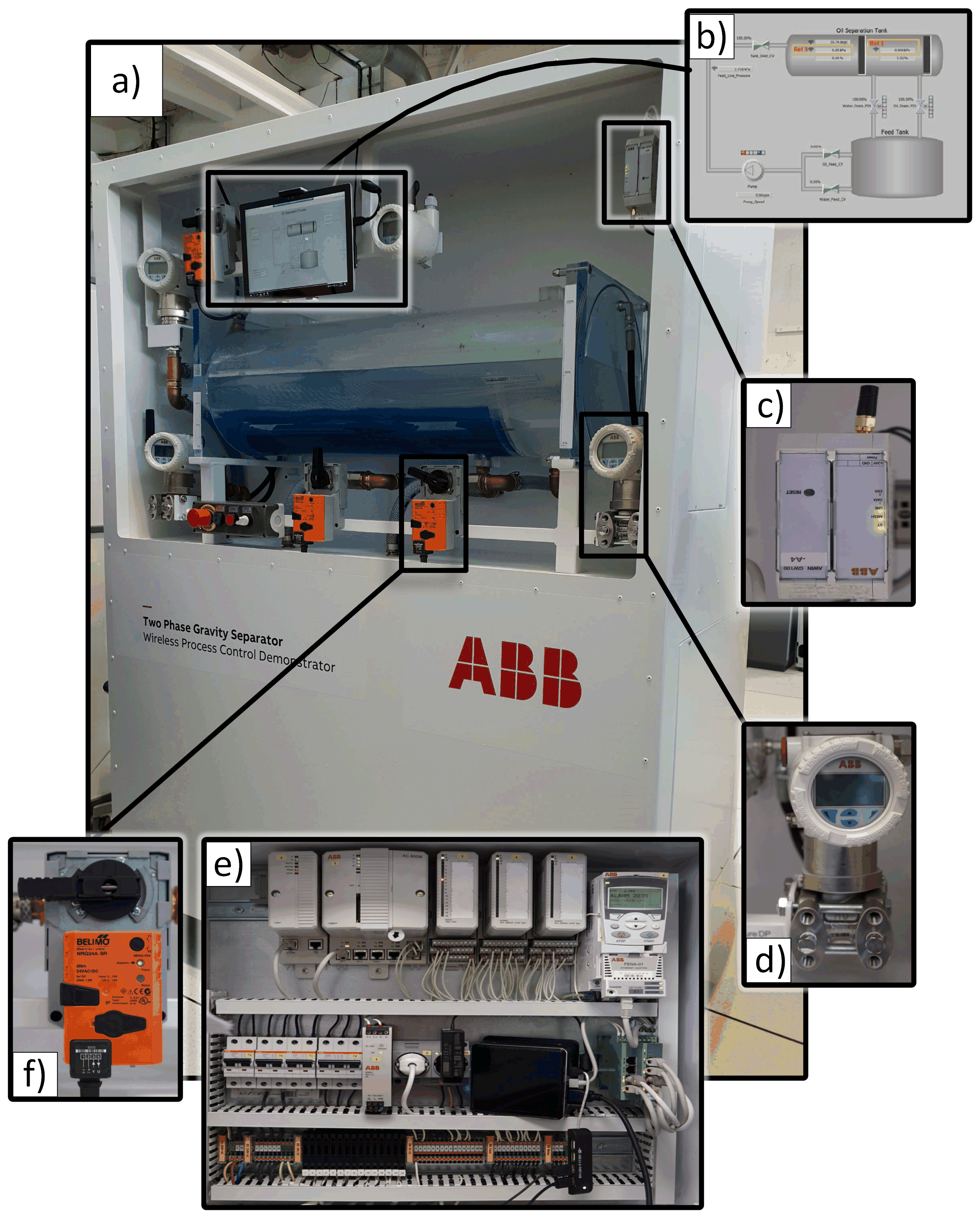}
    \caption{The Separator Testbed and its Components: a) Front view, b) Operator Workplace, c) WirelessHART gateway, d) Sensor node, e) Electrical cupboard with controller, f) Actuator node}
    \vspace{-5mm}
    \label{fig:testbed}
\end{figure}

The real Separator architecture is depicted in Figure~\ref{fig:testbed}.
Individual components of the Separator testbed are: sensor nodes, actuator nodes, controller, wireless network
and additional supporting components. These are described next. \vspace{2pt}\\
\textbf{Sensor Nodes} - The Separator has four industrial grade wireless sensors (depicted in Figure~\ref{fig:testbed}d) to measure the states of physical process. These are:
\begin{itemize}
    \item Two ABB DP-Style 266DSH differential pressure sensors (P2 and P3 in Figure~\ref{fig:separator})
    that are used to measure oil and water levels. The sensor P2 measures the water level in the left compartment of
    the tank based on the pressure difference between two liquids (oil and water) where water is at the bottom of the tank and oil on the top. The sensor P3 measures the oil level in the right compartment of the tank based on the pressure difference at the lowest (oil) and highest point (open-air) of the tank.  The differential pressure values measured
    by P2 and P3 are used as inputs for the PID controller.
    \item An ABB 266HSH High overload Pressure sensor (P1 in Figure~\ref{fig:separator}) that measures the absolute pressure
    in the feed line pipe. It is used as an input to the alarm system and the controller stops the system if the pressure
    in the pipe exceeds a threshold.
    \item An ABB TTF300-W WirelessHART Temperature sensor (T in Figure~\ref{fig:separator}) that measures the temperature of the oil in the tank. It
    is used as an input for the alarm system. The system stops its operation when the temperature of oil exceeds a threshold.
\end{itemize}
The differential pressure sensors require calibration before each run and it remains stable once operation has begun.\vspace{2pt}\\
\textbf{Actuator Nodes} - There are five Belimo NRQ24A-SR industrial grade valves in the system (depicted in Figure~\ref{fig:testbed}f).
These valves can go from fully open (100\%) to fully closed (0\%) in $9$ seconds. Two of these valves control the water
and oil inputs to the system (V1 and V2 in Figure~\ref{fig:separator}). The third valve controls the inlet level of the
mixture into the tank (V3 in Figure~\ref{fig:separator}). The other two valves are controlled by the PID controller to
ensure that the oil and water levels are at the set-points (LV1 and LV2 in Figure~\ref{fig:separator}).\vspace{2pt}\\
\textbf{Industrial Grade Controller} - 
The controller in  Figure~\ref{fig:testbed}e is an ABB AC800M programmable automation
with a CI867 Modbus TCP interface card 
The CI867 enables a Modbus TCP 
connection over Ethernet between the AC800M
controller, the AWIN GW100 WirelessHART gateway and the ACS355 motor drive.
In the Separator we tune two basic PID controllers to control the water and oil levels in the left and right compartments. 
The system tunes the levels of the oil first and then the levels of the water. The constant PID values are as
follows: for water $P = 1.4$, $I = 80$ and for oil $P = 2$, $I = 40$ and $D = 0$ in both cases. \vspace{2pt}\\
\textbf{Wireless Network} -
The wireless sensors communicate over WirelessHART protocol, an 
International Electrotechnical Commission (IEC) approved 
wireless communication protocol for wireless sensor networks. It is based on the Highway 
Addressable Remote Transducer (HART) protocol and uses 802.15.4 in the
2.4GHz ISM band. It forms a resilient self-organising network. Sensors 
nodes can find neighbours, detect failures, form communication routes and adjust these routes
based on sensor node failure. They form a mesh topology network where each node can 
act as a router for its neighbours. Each node connects directly with a minimum of two 
neighbours in order to provide this resiliency. 

WirelessHART organises the sensors so that they can transmit their readings to a central gateway, which forwards
its data over Ethernet to the controller. The WirelessHART gateway in the Separator is the AWIN GW100 
(depicted in Figure~\ref{fig:testbed}c). 
The gateway communicates using TCP
via the CI867 Modbus TCP interface to the AC800M controller.\vspace{2pt}\\
%
\textbf{Supporting Components} - There is an electrical cabinet on the rear of the Separator that 
houses the ACS355 Drive, the 24V DC power supply, the network switch and the Intel 
NUC PC (depicted in Figure~\ref{fig:testbed}e). It includes the distribution of the 220V AC and 24V DC to the
above equipment through four circuit breakers and a marshalling terminal for the IO signals.

The Separator has a Human Machine Interface (HMI) as a part of the Operator Workplace (depicted in 
Figure~\ref{fig:testbed}b).
It allows the user to see
information regarding the physical process, shows the state of the actuators and
the measured values from the sensors.
The user can intervene to stop the pump if deemed necessary, or to open/close a valve and even set the 
set-points at runtime.

\subsection{Safe Operation for Users of the System}
The last of our design requirements was that our CPS testbed be safe to use for research by 
students and researchers. To address this, The Separator was designed in compliance with ABB's 
process safety expertise. 
The Separator controller shuts down the system in the event that the water or oil level reach 80\%, to prevent an unsafe state. There is also an emergency switch under the separator tank which shuts the pump down when it is pressed. 
%
\section{Experimental Use Cases of The Separator}

In this section we present potential experimental use cases of The Separator. These include both the aspects of process 
control and system communication. We first present results showing the stability of the control process under stable 
operation. We then show how The Separator can adjust itself and maintain system stability when the oil level set point 
changes.
Finally, we show the robustness of the network and the controller to local radio
interference which causes slower communication. We evaluate CPS performance with four metrics that are defined as 
follows:
\begin{itemize}
    \item \textit{Latency} (in ms) - The average time required for a data packet to travel from the originating
    sensor to the controller.
    \item \textit{Path stability} (in \%) - The ratio of acknowledged packets to sent packets between two sensors or between a sensor and the gateway. 
    \item \textit{Overshoots} (OvSh in \%) - The percentage of the liquid (water or oil) above the set-point.
    This metric is important for ensuring safe behaviour of the system.
    \item \textit{Undershoots} (UnSh in \%) - The percentage of the liquid below the set-point after the first overshoot. This metric indicates the minimum level of liquid in the tank once the set-point has been reached.
\end{itemize}
The first two metrics describe the performance of the communication network, while the last two metrics describe
the performance of the control system.

\subsection{Use Case 1: Stable Process Operation}

\begin{figure}
    \centering
    \includegraphics[width=\linewidth]{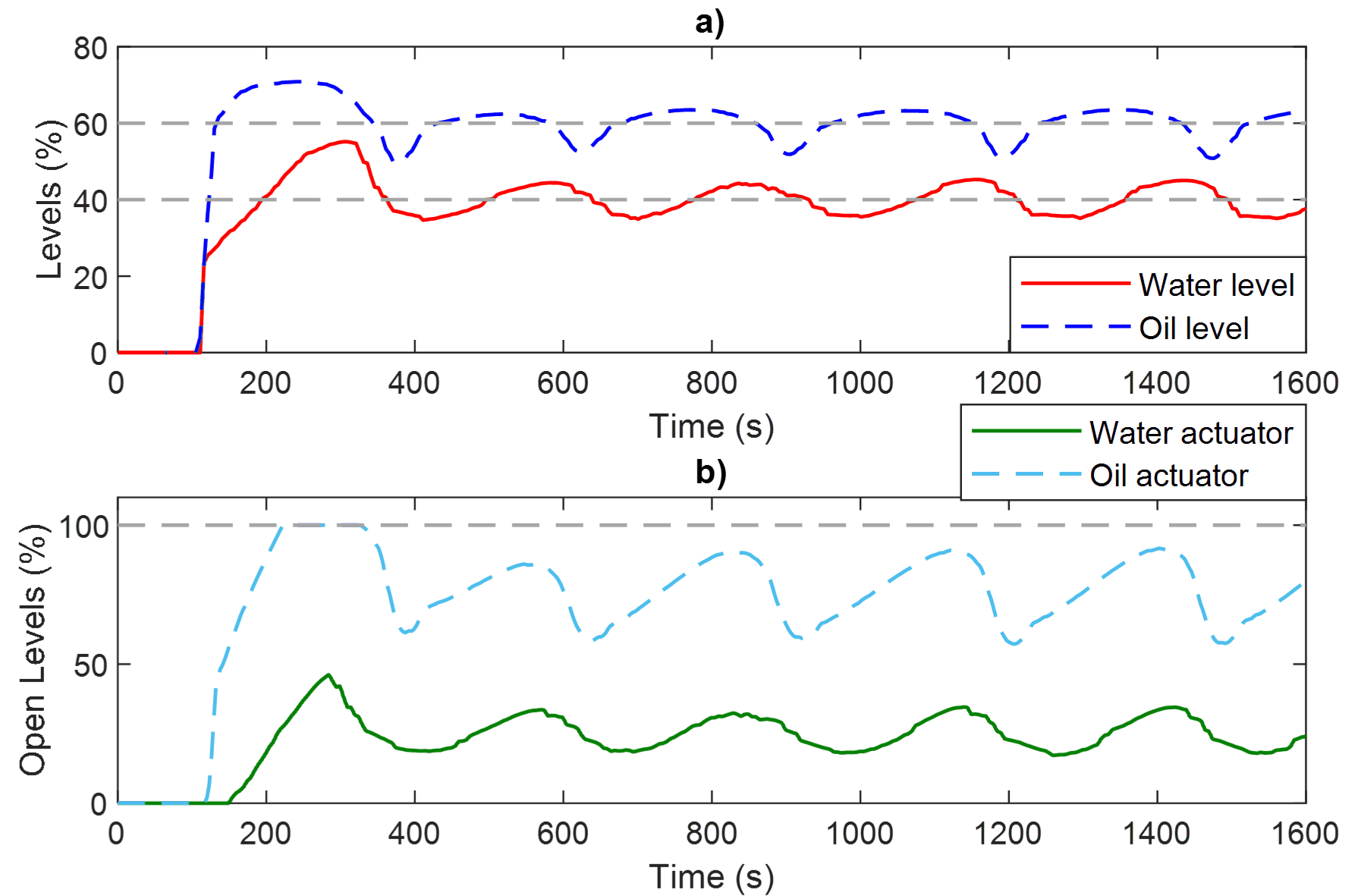}
    \caption{Stable process operation: a) Oil and water levels, b) Actuators open levels}
    \label{fig:oil_water}
\end{figure}

We first evaluate the stable operation of the testbed. We tune the PID controller and the inlet valves to
create a baseline experiment (as defined in Section~\ref{sec:pidtuning}). In this experiment, the goal of the system is to
maintain the levels of oil and water at the desired set-points of 60\% and 40\%, respectively. We run the
experiment five times. The results for overshoots
and undershoots in the left compartment (i.e. the water level) over four waves are shown in Table~\ref{tab:watererrors}. The results for 
latency and path stability are shown in Table~\ref{tab:networkstats}. We also present the measured oil and water levels and actuator levels 
for the left and right compartments in Figure~\ref{fig:oil_water}. The results in Table~\ref{tab:watererrors} show that the water level 
overshoots are almost indistinguishable for the second, third and four peaks. These peaks occur during stable operation. The first wave 
occurs as the system is stabilising, and so has larger overshoots, but has small standard deviation between experiments. These experiments 
show that The Separator can be used to perform reproducible experiments, with a very small standard deviation between the mean values 
during system stability.

\begin{table}[!t]
\caption{Stable process operation: Overshoots and Undershoots for water levels over 5 runs\vspace{-2mm}}
\begin{center}
\begin{tabularx}{0.44\textwidth}{l c c c c}
\toprule
 & \textbf{Wave 1} & \textbf{Wave 2} & \textbf{Wave 3} & \textbf{Wave 4} \\
\midrule
\textbf{OvSh (\%)} & 25.1704 & 10.2661 & 10.0240 & 10.5441 \\
\textbf{OvSh Std Dev(\%)} & 4.7141 & 1.3195 & 1.2979 & 0.8178 \\
\midrule
\textbf{UnSh (\%)} & 13.7424 & 12.1086 & 9.6345 & 8.8093 \\
\textbf{UnSh Std Dev(\%)} & 1.2343 & 0.7708 & 0.7244 & 1.0362 \\
\bottomrule
\end{tabularx} \vspace{-2mm}
\end{center}
\label{tab:watererrors}
\end{table}

\begin{table}[!t]
\caption{Stable process operation: Latency and Path stability over 5 runs \vspace{-2mm}}
\setlength{\tabcolsep}{4.4pt}
\begin{center}
\begin{tabularx}{0.49\textwidth}{l c c c c c}
\toprule
 & \textbf{Run 1} & \textbf{Run 2} & \textbf{Run 3} & \textbf{Run 4} & \textbf{Run 5} \\
\midrule
\textbf{Latency (ms)} & 0.35425 & 0.35925 & 0.35525 & 0.351 & 0.3515 \\
\textbf{Path Stability (\%)} & 99.5345 & 99.45125 & 99.5463 & 99.5075 & 99.3543 \\
\bottomrule
\end{tabularx} \vspace{-2mm}
\end{center}
\label{tab:networkstats}
\end{table}

\subsection{Use Case 2: Stable Operation with changing Oil Set-Point at Run-Time}

\begin{figure}
    \centering
    \includegraphics[width=0.9\linewidth]{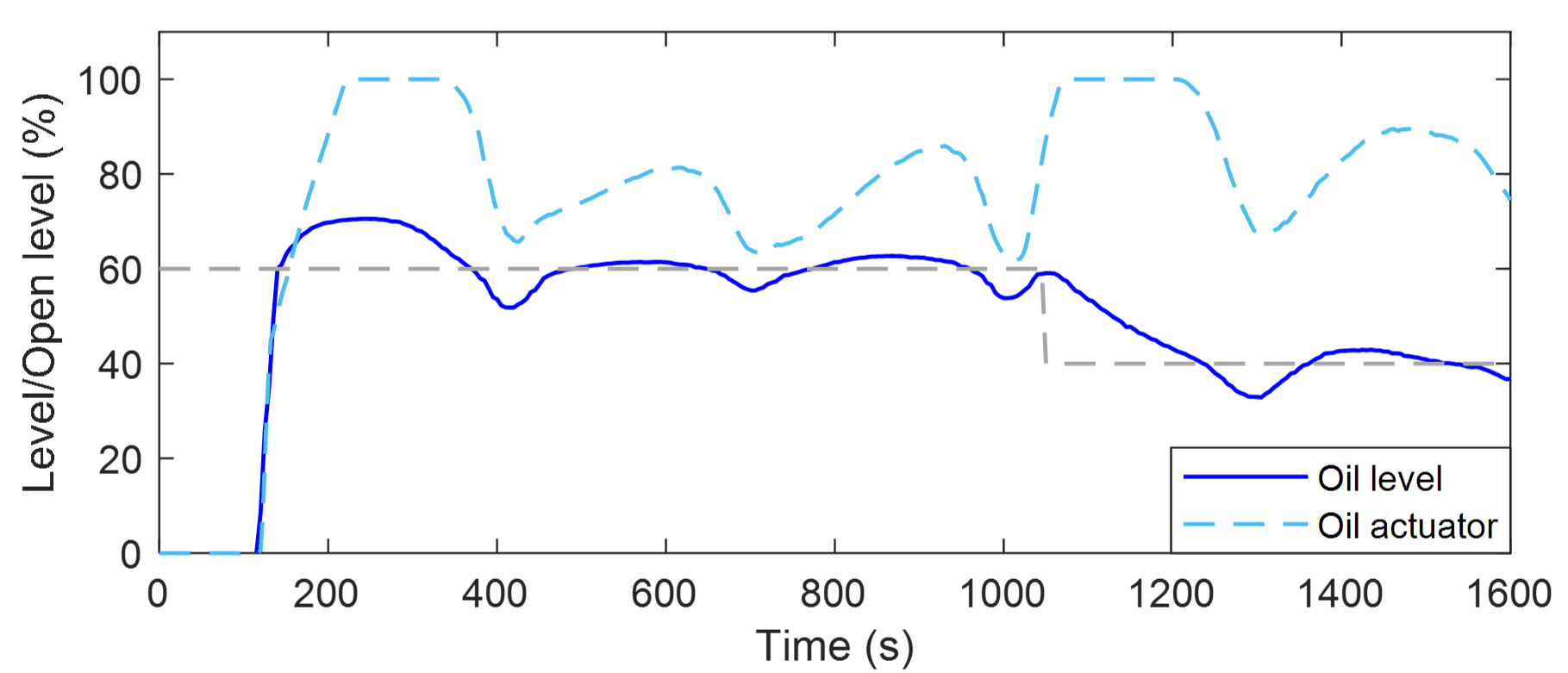}
    \caption{Oil levels and Actuator open levels for the set-point change}
    \label{fig:oil_adaptive}
\end{figure}

Use case 2 evaluates the capability of The Separator to maintain a stable system when the set-point changes. 
This allows us to evaluate the performance of the network and the 
controller for an application with requirements that change at run-time. We change the set-point for oil 
level from 60\% to 40\% after three overshoot peaks. We run the experiment five 
times and present the average overshoots, undershoots and the standard deviation after the set-point has been 
changed. The results are presented in Table~\ref{tab:oilerrors}. We also present the oil level 
and open-level of the actuators in Figure~\ref{fig:oil_adaptive}. The first undershoot occurs as the system is stabilising itself, but it is still small, and the average of the results has a standard deviation which is less than $1.2$\%. The system stabilises after the change very quickly, and by the subsequent overshoots are very small, and self-similar between runs with a standard deviation of less than $1$\%. These results show that The Separator can perform reproducible experiments with run-time configuration changes.

\begin{table}[!t]
\caption{Oil Errors, first one is undershoot other two overshoots for the set-point change from 60\% to 40\%}
\begin{center}
\begin{tabularx}{.36\textwidth}{l c c c}
\toprule
& \textbf{UnSh 1} & \textbf{OvSh 1} & \textbf{OvSh 2} \\
\midrule
\textbf{Error (\%)} & 14.5458 & 7.7229 & 6.1727 \\
\textbf{Error Std Dev(\%)} & 1.1636 & 0.7453 & 0.5893 \\
\bottomrule
\end{tabularx} \vspace{-2mm}
\end{center}
\label{tab:oilerrors}
\end{table}

\subsection{Use Case 3: Stable Operation in the Presence of Radio Interference}

The Separator contains sensors that communicate to a gateway via WirelessHART. This allows us to 
evaluate the robustness of the network and controller to adverse wireless communication conditions cause by
interference (potentially an attacker). To realise this, we used two USRP B210 software defined radios to
create white noise interference in the range of 802.15.4 channels 14, 15, 16, 23, 24, and 25. The affects are
shown in Table~\ref{tab:radiointerference}. We can see that the reliability, or percentage of messages received,
was 100\% for the entire trial. The link stability, or the number acknowledgements over the number of messages,
dropped due to the interference, and the packet delivery latency increased. The reliability of WirelessHART is unsurprising, it is a robust, multi-channel protocol that blacklists unreliable channels. Despite the robustness of WirelessHART, we can see
that the link stability and network performance were still affected by the interference. These results show that The Separator can also be used to evaluate system robustness in the presence of adverse communication conditions.
\begin{table}[!t]
\caption{Disruption of WirelessHART protocol}
\begin{center}
\begin{tabularx}{.45\textwidth}{l c c c}
\toprule
& \textbf{Reliability (\%)} & \textbf{Stability (\%)} & \textbf{Latency (ms)} \\
\midrule
\textbf{t = 0 min} & 100 & 99.38 & 0.296\\
\textbf{t = 7 min} & 100 & 88.26 & 0.452 \\
\textbf{t = 20 min} & 100 & 71.9 & 0.402 \\
\bottomrule
\end{tabularx} \vspace{-2mm}
\end{center}
\label{tab:radiointerference}
\end{table}

\section{Limitations and Future Research Directions}

Part of the research output of The Separator was the investigating what is needed for the research,
design, and engineering of reliable, safe CPSs. We tread the line between fidelity, with the use of industrial grade sensors, flexibility, with less robust research equipment. Things that we would consider differently would be:
\begin{enumerate}
    \item Industrial grade wireless transceivers can not run different communication protocols or add more sensors. For future iterations we would search for a flexible industrial level sensor platform.
    \item The current differential pressure sensors require calibration after every run/every time the tank is drained. In an industrial setting, this calibration is not time-consuming. The processes run for a long time. In research, multiple experiments are run. Sensors that need calibration every run add an overhead, and make it difficult to obtain repeatable results. We would search for industrial grade sensors less reliant on initial calibration.
\end{enumerate}

Finally, if we had unlimited funds we would choose to
build a larger-scale distributed system as many industrial systems are distributed over a number of
processes and over large distances. This would also allow for the deployment and evaluation of long-range communication protocols like LoRa.

A list of the future research opportunities with the current generation of The Separator, along with other associated
directions, is presented below: \vspace{2pt}\\
\textbf{Co-design of Communication and Control Techniques} - Explore different techniques to
co-design systems that integrate communication and control. Examine the use of \textit{formal
techniques} to model the protocol behaviour with the controller to both ensure that the
communication protocol maintains the properties required by the controller, and use \textit{run-time
verification} to check these properties during system operation.\vspace{2pt}\\
\textbf{Alternative Controller Schemes} - Compare of various modern
control schemes, such as \textit{event-triggered or self-triggered}, with currently used control schemes,
such as time triggered control. Experiment with other, non-conventional control approaches, such
as the use of \textit{machine learning} to learn 'black box controllers'. \vspace{2pt}\\
\textbf{Alternative Communication Schemes} - Use wired communication protocols like HART, Profibus, Fieldbus to communicate with the sensors. Integrate different communication technologies like WiFi, 802.15.4e and bluetooth low energy into The Separator.  \vspace{2pt}\\
%
\textbf{CPS Security Research} -  Assess potential vulnerabilities that the sensors and the 
communication may have to \textit{physical layer attacks} that aim to disrupt the sensor readings or the 
communication to the controller. \vspace{2pt}\\
\textbf{Teaching and Collaborations} - Use The Separator for teaching. It is the perfect testbed
illustrate several of the principles of CPSs, gives students \textit{hands-on experience} with
a real control system. The Separator will be available for \textit{collaborative projects} that 
allow the broader ICS community (both academia and industry) to contribute to our long-term goal of
transforming the ICS design process.


\section{Conclusion}

In this paper we presented The Separator, a CPS testbed whose inception and creation are part of the search
for approaches and techniques for the engineering of safe and stable CPS systems. The Separator is a result of 
the need to better understand the confluence of computer software and hardware systems, wireless communication and 
physical control processes. Preliminary experimental results demonstrate its high reliability, its use in 
performing reproducible experiments, and its ability to aid our understanding of how unreliable wireless 
communication and physical controllers can affect one another, and how to engineer this relationship to create 
stable and safe CPS.


\bibliographystyle{plain} 
\bibliography{references.bib}

\end{document}